\documentclass[11pt,a4paper]{article}

\usepackage{german}
\usepackage[ansinew]{inputenc}  
\usepackage[german,english]{babel}
\usepackage{latexsym,amsbsy,amssymb,amsmath,amsfonts}
\usepackage{epsfig}
\usepackage{euscript}
\usepackage{units}
\usepackage{fullpage}
\usepackage{pdflscape}
\usepackage{indent}
\usepackage{enumitem}
\usepackage{rotating}
\usepackage{xcolor}

\usepackage{rotating}
\usepackage{tikz}
\usetikzlibrary{shapes}
\usetikzlibrary{automata}
\usetikzlibrary{arrows}
\usetikzlibrary{calc}
\usepackage[colorlinks,linkcolor={red!75!black},citecolor={blue!75!black},urlcolor={blue!80!black}]{hyperref}

\newcommand{\opstyle}[1]{\mathrm{#1}}

\newcommand{\redstyle}[1]{\mathnormal{#1}}

\newcommand{\mr}{\:\!}
\newcommand{\ml}{\:\:\!\!\!}

\newcommand{\cP}{\opstyle{P}}
\newcommand{\cNP}{\opstyle{NP}}
\newcommand{\ccoNP}{\opstyle{coNP}}
\newcommand{\cUP}{\opstyle{UP}}

\newcommand{\FP}{\opstyle{FP}}

\newcommand{\Pol}{\opstyle{Pol}}

\newcommand{\calT}{{\cal T}}

\newcommand{\reduction}[3][]{%
    \redstyle{\le_{\mathrm{#3}}^{\mathrm{#2}#1}}}
\newcommand{\polyreduction}[2][]{\reduction[#1]{p}{#2}}
\newcommand{\redm}[1][]{\polyreduction[#1]{m}}

\newtheorem{dummytheorem}{Dummy-Theorem}[section]
\newcommand{\proofendsign}{$\Box$} 
\newtheorem{definition}[dummytheorem]{Definition}
\newtheorem{lemma}[dummytheorem]{Lemma}
\newtheorem{theorem}[dummytheorem]{Theorem}

\newtheorem{proposition}[dummytheorem]{Proposition}
\newtheorem{corollary}[dummytheorem]{Corollary}

\newtheorem{claim}[dummytheorem]{Claim}

\newenvironment{proof}{{\noindent \bf Proof }}{%
    {\hspace*{\fill}\proofendsign\par\bigskip}}
\newenvironment{beweis}{{\noindent \bf Beweis }}{%
    {\hspace*{\fill}\proofendsign\par\bigskip}}

\newcommand{\lqq}{\lq\lq}   
\newcommand{\rqq}{\rq\rq}   
\newcommand{\elqq}{\lqq}    
\newcommand{\erqq}{\rqq}    

\newcommand{\eqq}[1]{\elqq #1\erqq}

\newcommand{\isdefinedl}{\mathop{=}\limits^{\mbox{%
    \raisebox{-0.15ex}[0ex][0ex]{$\scriptscriptstyle df$}}}}

\newcommand{\isdefined}{\isdefinedl}

\newcommand{\oli}[1]{\overline{#1}}

\newcommand{\az}{{\Sigma}} 
\newcommand{\sow}{{\az^*}}

\newcommand{\rem}[1]{}

\newcommand{\tn}[1]{\textnormal{#1}}    

\newsavebox{\einzugbox}

\newenvironment{lind}[1]{\begin{indentation}{#1}{0mm}}{\end{indentation}}

\newlength{\chrlengtha}
\newlength{\chrlengthb}

\newcommand{\algob}{\begin{lind}{5mm}\fns}
\newcommand{\algoe}{\end{lind}}

\newcommand{\N}{\mathbb{N}}
\newcommand{\Z}{\mathbb{Z}}

\newcommand{\pairing}[1]{\langle #1 \rangle}

\newcommand{\pr}{\mathrm{pr}}   

\newcommand{\fns}{\footnotesize}

\begin{document}
\selectlanguage{english}

\def\sqsubsetneq{\mathrel{\sqsubseteq\kern-0.92em\raise-0.15em\hbox{\rotatebox{313}{\scalebox{1.1}[0.75]{\(\shortmid\)}}}\scalebox{0.3}[1]{\ }}}
\def\sqsupsetneq{\mathrel{\sqsupseteq\kern-0.92em\raise-0.15em\hbox{\rotatebox{313}{\scalebox{1.1}[0.75]{\(\shortmid\)}}}\scalebox{0.3}[1]{\ }}}

\newcommand{\io}{\mathrm{io\tn{-}}}

\newcommand{\ioeq}{\mathop{\mbox{$\mathop{=}\limits^{\mathrm{\scriptscriptstyle io}}$}}}

\newcommand{\notioeq}{\mathop{\mbox{$\mathop{=}\limits^{\mathrm{\scriptscriptstyle \mr i\ml .\ml o\ml .}}\!\!\!\!\!\raisebox{0.3ex}[0ex][0ex]{$/$}\mr\mr\mr$}}}

\newcommand{\ioin}{\mathop{\mbox{$\mathop{\in}\limits^{\mbox{\raisebox{-0.2ex}[0ex][0ex]{$\mathrm{\scriptscriptstyle \mr\mr i\ml .\ml o\ml .}$}}}$}}}

\newcommand{\notioin}{\mathop{\mbox{$\mathop{\in}\limits^{\mbox{\raisebox{-0.2ex}[0ex][0ex]{$\mathrm{\scriptscriptstyle \mr\mr i\ml .\ml o\ml .}$}}}\!\!\!\!\!\ml\raisebox{0.35ex}[0ex][0ex]{$/$}\mr\mr\mr$}}}

\newcommand{\iosub}{\mathop{\mbox{$\mathop{\subseteq}\limits^{\mbox{\raisebox{-0.2ex}[0ex][0ex]{$\mathrm{\scriptscriptstyle \mr\mr i\ml .\ml o\ml .}$}}}$}}}

\newcommand{\notiosub}{\mathop{\mbox{$\mathop{\subseteq}\limits^{\mbox{\raisebox{-0.2ex}[0ex][0ex]{$\mathrm{\scriptscriptstyle \mr\mr i\ml .\ml o\ml .}$}}}\!\!\!\!\!\ml\raisebox{0.35ex}[0ex][0ex]{$/$}\mr\mr\mr$}}}

\newcommand{\aeeq}{\mathop{\mbox{$\mathop{=}\limits^{\mathrm{\scriptscriptstyle \mr a\ml .\ml e\ml .}}$}}}

\newcommand{\notaeeq}{\mathop{\mbox{$\mathop{=}\limits^{\mathrm{\scriptscriptstyle \mr a\ml .\ml e\ml .}}\!\!\!\!\!\raisebox{0.3ex}[0ex][0ex]{$/$}\mr\mr\mr$}}}

\newcommand{\aein}{\mathop{\mbox{$\mathop{\in}\limits^{\mbox{\raisebox{-0.2ex}[0ex][0ex]{$\mathrm{\scriptscriptstyle \mr\mr a\ml .\ml e\ml .}$}}}$}}}

\newcommand{\notaein}{\mathop{\mbox{$\mathop{\in}\limits^{\mbox{\raisebox{-0.2ex}[0ex][0ex]{$\mathrm{\scriptscriptstyle \mr\mr a\ml .\ml e\ml .}$}}}\!\!\!\!\!\ml\raisebox{0.35ex}[0ex][0ex]{$/$}\mr\mr\mr$}}}

\newcommand{\aesub}{\mathop{\mbox{$\mathop{\subseteq}\limits^{\mbox{\raisebox{-0.2ex}[0ex][0ex]{$\mathrm{\scriptscriptstyle \mr\mr a\ml .\ml e\ml .}$}}}$}}}

\newcommand{\notaesub}{\mathop{\mbox{$\mathop{\subseteq}\limits^{\mbox{\raisebox{-0.2ex}[0ex][0ex]{$\mathrm{\scriptscriptstyle \mr\mr a\ml .\ml e\ml .}$}}}\!\!\!\!\!\ml\raisebox{0.35ex}[0ex][0ex]{$/$}\mr\mr\mr$}}}

\newcommand{\redmiopoly}[1][]{\reduction[#1]{\io p/poly}{m}}
\newcommand{\psim}[1][]{\le^{\mathrm{p}#1}}

\newcommand{\NPCm}{\tn{NPC}^{\tn{p}}_{\tn{m}}}
\newcommand{\NPCO}{\tn{NPC}^{\tn{p,}O}_{\tn{m}}}
\newcommand{\NPCT}{\tn{NPC}^{\tn{p}}_{\tn{T}}}
\newcommand{\NPCmpoly}{\tn{NPC}^{\tn{p/poly}}_{\tn{m}}}
\newcommand{\NPCmiopoly}{\tn{NPC}^{\tn{io-p/poly}}_{\tn{m}}}
\newcommand{\NPCtwott}{{\rm NPC}^{\rm p}_{\text{\rm 2-tt}}}
\newcommand{\NPCli}{\text{\rm NPC}_{\rm m, li}^{\rm p}}
\renewcommand{\Pol}{\tn{Pol}}
\newcommand{\Pad}{\tn{Pad}}
\newcommand{\Hom}{\tn{Hom}}
\newcommand{\Hm}{\tn{H}^{\tn{p}}_{\tn{m}}}
\newcommand{\Hmli}{\tn{H}^{\tn{p}}_{\tn{m,li}}}
\newcommand{\Hone}{\tn{H}^{\tn{p}}_{\tn{1}}}
\newcommand{\HT}{\tn{H}^{\tn{p}}_{\tn{T}}}
\newcommand{\HsnT}{\tn{H}^{\tn{p}}_{\tn{snT}}}
\newcommand{\Hmpoly}{\tn{H}^{\tn{p/poly}}_{\tn{m}}}
\newcommand{\Hmiopoly}{\tn{H}^{\tn{io\tn{-}p/poly}}_{\tn{m}}}

\renewcommand{\P}{\mathbb{P}}
\newcommand{\Podd}{\P^{\ge 3}}

\newcommand{\ran}{\tn{ran}}

\newcommand{\DisjcoNP}{\mathrm{DisjCoNP}}

\title{P-Optimal Proof Systems for Each coNP-Complete Set and \\no Complete 
       Problems in $\cNP\cap\ccoNP$ Relative to an Oracle}

\author{Titus Dose\\Julius-Maximilians-Universität Würzburg}
\maketitle

\begin{abstract}
    We build on a working program initiated by Pudlák \cite{pud17}
    and construct an oracle relative to which each $\ccoNP$-complete set
    has $\cP$-optimal proof systems and $\cNP\cap\ccoNP$ does not have
    complete problems.
\end{abstract}

\section{Introduction}
The main motivation for the present paper is an article by Pudlák \cite{pud17} 
who lists several major conjectures in the field
of proof complexity and discusses their relations. 
Among others, Pudl{\'a}k conjectures the following assertions (note that within the present
paper all reductions are polynomial-time-bounded):
\begin{itemize}
	\item   $\mathsf{CON}$ (resp., $\mathsf{SAT}$): $\ccoNP$ (resp., $\cNP$) does not contain many-one complete
            sets that have P-optimal proof systems 
    \item   $\mathsf{CON}^{\mathsf{N}}$: $\ccoNP$ does not contain many-one complete
            sets that have optimal proof systems,\\
            (note that $\mathsf{CON}^{\mathsf{N}}$ is the non-uniform version of
            $\mathsf{CON}$)
    \item   $\mathsf{DisjNP}$ (resp., $\mathsf{DisjCoNP}$): The class of all disjoint $\cNP$-pairs 
            (resp., $\ccoNP$-pairs) does not have many-one complete elements,
    \item   $\mathsf{TFNP}$: The class of all total polynomial search problems does not have complete elements,
    \item   $\mathsf{NP}\cap\mathsf{coNP}$ (resp., $\mathsf{UP}$): $\cNP\cap\ccoNP$ (resp., $\cUP$, 
            the class of problems accepted by $\cNP$ machines with at most one accepting path
            for each input) does not have many-one complete elements.
\end{itemize}
Pudlák asks for oracles separating corresponding relativized conjectures.
Recently there has been made some progress in this working program
\cite{kha19,dg19,dos19,dos19b} which is documented by the following figure 
representing the current state of the art.
\newcommand{\CCON}{\mathsf{CON}}
\newcommand{\CCONN}{\mathsf{CON}^\mathsf{N}}
\newcommand{\CDisjNP}{\mathsf{DisjNP}}
\newcommand{\CUP}{\mathsf{UP}}
\newcommand{\CRFN}{\mathsf{RFN}_1}
\newcommand{\CPNP}{\mathsf{P}\ne\mathsf{NP}}
\newcommand{\CNPcoNP}{\mathsf{NP}\cap\mathsf{coNP}}
\newcommand{\CSAT}{\mathsf{SAT}}
\newcommand{\CTFNP}{\mathsf{TFNP}}
\newcommand{\CDisjCoNP}{\mathsf{DisjCoNP}}

\begin{center}
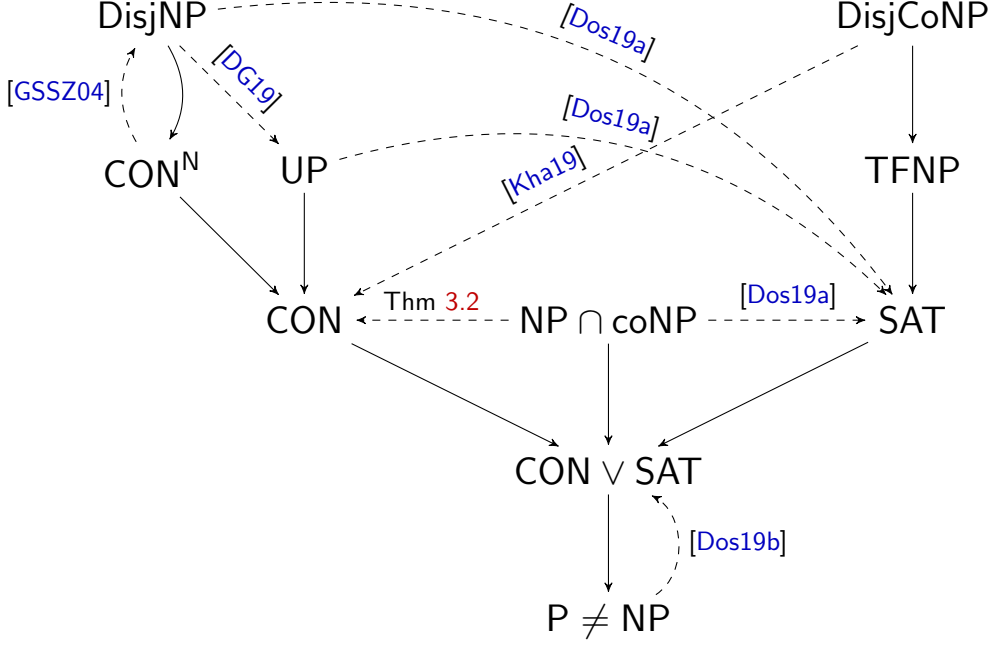
\begin{figure}[ht]
\begin{tikzpicture}[->,>=stealth',initial text={},shorten >=1pt,auto,node distance=2cm,
  thin,main node/.style={draw=none,font=\sffamily\Large}, thin]

  \node[main node] (1)  {$\CDisjNP$};
  \node[main node] (2) [below of=1] {$\CCONN$};
  \node[main node] (3) [right  of=2] {$\CUP$};
  \node[main node] (4) [ below of=3] {$\CCON$};
\node[main node] (12) [right of=4]{};
\node[ main node] (14) [ below of=4] {};
\node[ main node] (15) [ right of=14] {};
	\node[ main node] (5) [ right of=15] {$\CCON\vee\CSAT$};
  \node[ main node] (6) [ below  of=5] {$\CPNP$};
  \node[ main node] (7) [ right of=12] {$\CNPcoNP$};
\node[main node] (13) [right of=7]{};
	\node[ main node] (8) [ right of=13] {$\CSAT$}; 
	\node[ main node] (9) [ above of=8] {$\CTFNP$};
	\node[ main node] (10) [ above of=9] {$\CDisjCoNP$};

  \path[every node/.style={font=\sffamily\small, color  = black}]
    (1) edge[bend left] node[left] {} (2)
     (2)   edge [] node [] {} (4)
  (3) edge node [] {} (4)
  (4) edge node [] {} (5)
	(5) edge node [] {} (6)
	(7) edge node [] {} (5)
	(8) edge node [] {} (5)
    (9) edge node [] {} (8)
    (10) edge [] node  {} (9);
    
    \path[every node/.style={font=\sffamily\small, color  = black},dashed]
    (1) edge node [above,sloped] {\cite{dg19}} (3)
    (7) edge node [sloped,above] {Thm~\ref{theorem_0917240914}} (4)
    (1) edge [bend left] node [above,sloped] {\cite{dos19}} (8)
    (3) edge [bend left] node [sloped,above] {\cite{dos19}\phantom{........}} (8)
    (7) edge [] node [above] {\cite{dos19}} (8)
    (10) edge [] node [above,sloped] {\cite{kha19}\phantom{...............}} (4)
    (2) edge[bend left] node[left]{\cite{gssz04}}(1)
    (6) edge[bend right=60] node[right] {\cite{dos19b}}(5);
\end{tikzpicture}
\caption{\label{fig_1047120471}
Solid arrows mean implications. All implications occurring in
the graphic have relativizable proofs. A dashed arrow from
one conjecture $\mathsf{A}$ to another conjecture $\mathsf{B}$ means that
there is an oracle $X$ against the implication $\mathsf{A}\Rightarrow
\mathsf{B}$, i.e., relative to $X$, it holds $\mathsf{A}\wedge\neg
\mathsf{B}$.
\newline Pudl{\'a}k \cite{pud17} also defines the conjecture $\CRFN$ and lists it between $\CCON\vee\CSAT$ and
$\CPNP$, i.e., $\CCON\vee\CSAT\Rightarrow \CRFN\Rightarrow \CPNP$. Khaniki \cite{kha19} even shows $\CCON\vee\CSAT\Leftrightarrow \CRFN$, which is why we omit $\CRFN$ in the figure. 
For a definition of $\CRFN$ we refer to \cite{pud17}.
}
\end{figure}
\end{center}

In the figure Thm~\ref{theorem_0917240914} denotes the result of the present paper.
It shows that there is no relativizable proof for the implication
$\CNPcoNP \Rightarrow \CCON$. So the conjectures $\CNPcoNP$ and
$\CCON$ cannot be shown equivalent with relativizable proofs.

\section{Preliminaries} \label{sec_prelim}
Throughout this paper let $\Sigma$ be the alphabet $\{0,1\}$.
We denote the length of a word
$w\in\sow$ by $|w|$.
Let $\sow^{\prec n} = \{w \in \sow ~|~ |w| \prec n\}$ for $\prec\in
\{\le,<,=,>,\ge\}$.
The empty word is denoted by $\varepsilon$ and
the $i$-th letter of a word $w$ for $0 \le i < |w|$ is denoted by $w(i)$, i.e.,
$w = w(0) w(1) \cdots w(|w|-1)$.
For $k \le |w|$ let $\pr_k(w) = w(0) \cdots w(k-1)$
be the length $k$ prefix of $w$. A word $v$ is a prefix of $w$ if
there exists $k\le |w|$ such that $v = \pr_k(w)$.
If $v$ is a prefix of $w$, then we write $v \sqsubseteq w$ or $w\sqsupseteq v$.
If $v\sqsubseteq w$ and $|v| < |w|$, then we write $v\sqsubsetneq w$ or $w\sqsupsetneq v$.
For each finite set $Y \subseteq \sow$, let
$\ell(Y) \isdefined \sum_{w \in Y} |w|$.

Given two sets 
$A$ and $B$, $A-B$ denotes the set difference
between $A$ and $B$, i.e., $A-B = \{a\in A\mid a\notin B\}$. 
The complement of a set $A$ relative to the universe $U$ 
is denoted by $\oli{A}= U-A$. The universe will always be
apparent from the context.  Furthermore, the symmetric difference
is denoted by $\triangle$, i.e., $A\triangle B = (A-B)\cup (B-A)$ 
for arbitrary sets $A$ and $B$.

$\mathbb{Z}$ denotes the set of integers,
$\N$ denotes the set of natural numbers, and $\N^+ = \N-\{0\}$.
The set of primes is denoted by $\mathbb{P} = \{2,3,5,\ldots\}$
and $\Podd$ denotes the set $\mathbb{P}-\{2\}$.

We identify $\sow$ with $\N$
via the polynomial-time computable, polynomial-time invertible bijection
$w \mapsto \sum_{i<|w|} (1+w(i)) 2^{|w|-1-i}$,
which is a variant of the dyadic encoding.
Hence, notations, relations, and operations for $\sow$
are transferred to $\N$ and vice versa.
In particular, $|n|$ denotes the length of $n \in \N$.
We eliminate the ambiguity of the expressions $0^i$ and $1^i$
by always interpreting them over $\sow$.


Let $\langle \cdot \rangle : \bigcup_{i \ge 0} \N^i \rightarrow \N$
be an injective, polynomial-time computable, polynomial-time invertible
pairing function such that
$|\pairing{u_1, \ldots, u_n}| = 2(|u_1| + \cdots + |u_n| + n)$.

The domain and range of a function $t$ are denoted by
$\tn{dom}(t)$ and $\ran(t)$, respectively.

$\FP$, $\cP$, and $\cNP$ denote standard complexity classes \cite{pap94}.
Define $\opstyle{co\mathcal{C}} = \{A\subseteq\Sigma^*\mid \oli{A} \in\mathcal{C}\}$
for a class $\mathcal{C}$. 

We also consider these complexity classes in the presence of
an oracle $D$ and denote the corresponding classes by $\FP^D$, $\cP^D$, and $\cNP^D$.
Moreover, we define $\opstyle{co\mathcal{C}}^D = \{A\subseteq\Sigma^*\mid \oli{A} \in\mathcal{C}^D\}$
for a class $\mathcal{C}$. 

Let $M$ be a Turing machine. $M^D(x)$ denotes the 
computation of $M$ on input $x$ with $D$ as an oracle. For an
arbitrary oracle $D$ we let $L(M^D) = \{ x ~|~ M^D(x) \tn{ accepts}\}$,
where as usual in case $M$ is nondeterministic, the computation $M^D(x)$ 
accepts if and only if it has at least one accepting path.

For a deterministic polynomial-time Turing transducer (i.e., a Turing machine computing
a function), depending on the context,
$F^D(x)$ either denotes the computation of $F$ on input $x$ with $D$ as an oracle
or the output of this computation.

\begin{definition}\label{definition_019470123712}
A sequence $(M_i)_{i\in\N^+}$ is called {\em standard enumeration} of
nondeterministic, polynomial-time oracle Turing machines,
if it has the following properties:
\begin{enumerate}
    \item All $M_i$ are nondeterministic, polynomial-time oracle Turing machines.
    \item For all oracles $D$ and all inputs $x$
    the computation $M_i^D(x)$ stops within $|x|^i + i$ steps.
    \item For every nondeterministic, polynomial-time oracle Turing machine $M$
    there exist infinitely many $i \in \N$ such that
    for all oracles $D$ it holds that $L(M^D) = L(M_i^D)$.
    \item There exists a nondeterministic, polynomial-time oracle Turing machine $M$
    such that for all oracles $D$ and all inputs $x$ it holds that $M^D(\langle i,x,0^{|x|^i+i} \rangle)$
    nondeterministically simulates the computation $M_i^D(x)$.
\end{enumerate}
Analogously we define standard enumerations of
deterministic, polynomial-time oracle Turing transducers.
\end{definition}
Throughout this paper, we fix some standard enumerations.
Let $M_1,M_2,\dots$ be a standard enumeration of nondeterministic
polynomial-time oracle Turing machines. Then for every oracle $D$, the sequence
$(M_i)_{i\in\N^+}$ represents an enumeration of the languages in $\cNP^D$, i.e.,
$\cNP^D = \{L(M_i^D)\mid i\in\N\}$. 
Let $F_1,F_2,\dots$ be a standard enumeration of polynomial time 
oracle Turing transducers. 

By the properties of standard enumerations, for each oracle $D$ the
problem
$$K^D = \{\pairing{0^i,0^t,x}\mid i,t,x\in\N, i>0, \tn{ and $M_i^D(x)$ accepts within $t$ steps}  \}$$
is $\cNP^D$-complete (in particular it is in $\cNP^D$) and consequently, $\oli{K^D}$ is $\ccoNP^D$-complete.

In the present article we only use polynomial-time-bounded many-one reductions.
Let $D$ be an oracle. For problems $A,B\subseteq\sow$ we write $A\redm B$ (resp.,
$A\redm[,D] B$) if there exists $f\in\FP$ (resp., $f\in\FP^D$) with 
$\forall_{x\in\az^*} \big(x\in A\Leftrightarrow f(x)\in B$\big). In this case we say that $A$
is polynomially many-one reducible to $B$.

\begin{definition}[\cite{cr79}]
    A function $f \in \FP$ is called {\em proof system} for the set $\ran(f)$.
    For $f,g \in \FP$ we say that {\em $f$ is simulated by $g$} (resp., 
    {\em $f$ is $\cP$-simulated by $g$}) denoted by $f\le g$ (resp., $f \psim g$),
    if there exists a function $\pi$ (resp., a function $\pi \in \FP$)
    and a polynomial $p$ such that $|\pi(x)|\le p(|x|)$ and $g(\pi(x)) = f(x)$ for all $x$.
    A function $g \in \FP$ is {\em optimal} (resp., {\em $\cP$-optimal}),
    if $f\le g$ (resp., $f \psim g$) for all $f \in \FP$ with $\tn{ran(f)}=\tn{ran(g)}$.
    Corresponding relativized notions are obtained by using
    $\cP^D$, $\FP^D$, and $\psim[,D]$ in the definitions above.
\end{definition}The following proposition states the relativized version of a result by
K{\"o}bler, Messner, and Tor\'an \cite{kmt03},
which they show with a relativizable proof.
\begin{proposition}[\cite{kmt03}] \label{propo_pps_oracle}
    For every oracle $D$,
    if $A$ has a $\cP^D$-optimal (resp., optimal) proof system
    and $B \redm[,D]\! A$,
    then $B$ has a $\cP^D$-optimal (resp., optimal) proof system.
\end{proposition}

\begin{corollary} \label{coro_pps_oracle}
    For every oracle $D$,
    if there exists a $\redm[,D]\!$-complete $A \in \ccoNP^D$
    that has a $\cP^D$-optimal (resp., optimal) proof system,
    then all sets in $\ccoNP^D$
    have $\cP^D$-optimal (resp., optimal) proof systems.
\end{corollary}

\medskip
Let us introduce some (partially quite specific) notations that are designed
for the construction of oracles \cite{dg19}.
The support $\tn{supp}(t)$ of a real-valued function $t$ 
is the subset of the domain that 
consists of all values that $t$ does not map to 0.
We say that a partial function $t$ is injective on its support if 
$t(i,j) = t(i',j')$ for $(i,j),(i',j') \in \tn{supp}(t)$ implies $(i,j) = (i',j')$. 
If a partial function $t$ is not defined at point $x$,
then $t \cup \{x \mapsto y\}$ denotes the extension of $t$
that at $x$ has value $y$.

If $A$ is a set, then $A(x)$ denotes the characteristic function at point $x$,
i.e., $A(x)$ is $1$ if $x \in A$, and $0$ otherwise.
An oracle $D \subseteq \N$ is identified with its characteristic sequence
$D(0) D(1) \cdots$, which is an $\omega$-word.
In this way, $D(i)$ denotes both, the characteristic function at point $i$ and
the $i$-th letter of the characteristic sequence, which are the same.
A finite word $w$ describes an oracle that is partially defined, i.e.,
only defined for natural numbers $x<|w|$.
We can use $w$ instead of the set $\{i ~|~ w(i)=1 \}$ and
write for example $A = w \cup B$, where $A$ and $B$ are sets.
For nondeterministic oracle Turing machines $M$
we use the following phrases:
a computation $M^w(x)$ {\em definitely accepts},
if it contains a path that accepts and all queries 
on this path are $<|w|$.
A computation $M^w(x)$ {\em definitely rejects},
if all paths reject and all queries are $<|w|$.

For a nondeterministic Turing machine $M$ we say 
that the computation $M^w(x)$ {\em is defined},
if it definitely accepts or definitely rejects.
For a polynomial-time oracle transducer $F$, the computation $F^w(x)$ {\em is defined}
if all queries are $<|w|$.

\section{Oracle Construction}
The following lemma is a slightly adapted variant of a result from
\cite{dg19}.
\begin{lemma}\label{84189612339821692813}
For all $y\le |w|$ and all $v\sqsupseteq w$ it holds $(y\in K^v \Leftrightarrow y\in K^w)$.
\end{lemma}
\begin{proof} We may assume $y = \pairing{0^i,0^t,x}$ for suitable $i\in\N^+$ and $t,x\in\N$,
        since otherwise, $y\notin K^w$ and $y\notin K^v$.
        For each $q$ that is queried within the first $t$ steps
        of $M_i^w(x)$ or $M_i^v(x)$ it holds that
        $|q| \le t < |y|$ and thus, $q < y$.
        Hence, these queries are answered the same way relative to $w$ and $v$,
        showing that $M_i^w(x)$ accepts within $t$ steps if and only if $M_i^v(x)$ accepts
        within $t$ steps.
\end{proof}

\begin{theorem}\label{theorem_0917240914}
There exists an oracle $O$ such that
the following statements hold:
\begin{itemize}
	\item $\cNP^O\cap \ccoNP^O$ does not have $\redm[,O]$-complete problems.
    \item $\oli{K}^O$ has $\cP^O$-optimal proof systems.
\end{itemize}
\end{theorem}
\begin{proof}{\bf of Theorem~\ref{theorem_0917240914}}
Let $D$ be a (possibly partial) oracle and $p\in\Podd$.
We define
\begin{eqnarray*}
A_p^D &:=& \{0^{p^k} \mid k\in\N^+, \exists_{x\in
\az^{p^k}} x\in D\tn{ and $x$ odd}\} \cup \oli{
\{0^{p^k}\mid k\in\N^+\}}\\
B_p^D &:=& \{0^{p^k} \mid k\in\N^+, \exists_{x\in
\az^{p^k}} x\in D\tn{ and $x$ even}\}
\end{eqnarray*} 
Note that $A_p^D,B_p^D\in\cNP^D$ and $A_p^D = \oli{
B_p^D}$ if $|\az^{p^k}\cap D| = 1$ for each $k\in\N^+$. 
In that case $A_p^D\in\cNP^D\cap\ccoNP^D$.

For the sake of simplicity, let us call a pair $(M_i,M_j)$ an $\cNP^D\cap\ccoNP^D$-machine
if $L(M_i^D) = \oli{L(M_j^D)}$. Note that throughout this proof we sometimes omit the oracles in
the superscript, e.g., we write $\cNP$ or
$A_p$ instead of $\cNP^D$ or $A_p^D$. However, 
we do not do that in the ``actual'' proof but only when 
explaining ideas in a loose way
in order to give the reader the intuition behind the 
occasionally very technical arguments.

{\em Preview of construction.} We sketch some very basic ideas of our construction.
\begin{enumerate}
	\item For all $i>0$ we try to ensure that $F_i$ is not a proof system for $\oli{K}$
    relative to the final oracle. If this is possible, we do not have to consider $F_i$ anymore.
    If it is not possible, then $F_i$ inherently is a proof system for $\oli{K}$. In that case
    we start to encode the values of $F_i$ into the oracle. This way we easily obtain a
    $\cP$-optimal proof system for $\oli{K}$ in the end.
    Note that it is crucial that we allow to also encode values of functions $F_j$ into the oracle
    before we try ---as described above--- to make sure 
    that these functions are not proof systems for $\oli{K}$. 
    Hence, the final oracle also contains
    encodings of values of functions that are not proof systems for $\oli{K}$.
    \item Similarly, for each pair $(i,j)$ with $i\ne j$ we first try to make sure that $(M_i,M_j)$ 
    is not a $\cNP\cap\ccoNP$-machine. If this is not possible, then $(M_i,M_j)$ inherently is an $\cNP
    \cap\ccoNP$-machine. In this case we choose a prime $p$ and ensure in
    the further construction that $A_p = \oli{B_p}$, i.e., $A_p\in\cNP\cap\ccoNP$. Moreover,
    we diagonalize against all $\FP$-functions $F_r$ in order to make sure that $F_r$ does not
    reduce $A_p$ to $L(M_i)$.
\end{enumerate}

\bigskip For $i\in\N^+$ and $x,y\in\N$ 
we write $c(i,x,y) := \pairing{0^i, 0^{|x|^i + i}, 0^{|x|^i + i},x,y,y}$. 
Note that $|c(i,x,y)|$ is even and by the properties
of the pairing function $\pairing{\cdot}$,
\begin{align}\label{eq_140712041003}
\forall_{i\in\N^+, x,y\in\N}\;|c(i,x,y)| > 4\cdot \max( |x|^i + i,|y|).
\end{align}
\begin{claim} \label{claim_7378194873}
        Let $w\in\az^*$ be an oracle, $i\in\N^+$, and $x,y\in\N$ such that
        $c(i,x,y)\le |w|$. Then the following holds.
        \begin{enumerate}
            \item $F_i^w(x)$ is defined and $F_i^w(x)<|w|$.
            \item $(F_i^w(x)\in K^w\Leftrightarrow F_i^w(x)\in K^v)$
            for all $v \sqsupseteq w$.
        \end{enumerate}
    \end{claim}
\begin{proof}
As the running time of $F_i^w(x)$ is bounded by 
$|x|^i + i < |c(i,x,y)| < c(i,x,y) \le |w|$,
the computation $F_i^w(x)$ is defined and its output 
is less than $|w|$. Hence, 1 holds.
Consider 2. It suffices to show that 
$K^v(q) = K^w(q)$ for all $q < |w|$ and all $v\sqsupseteq
w$. This holds by Lemma~\ref{84189612339821692813}.
\end{proof}

During the construction we maintain a growing collection 
of requirements that is represented by a partial function belonging to the set
$$\calT = \Big\{t:(\N^+)^2\to\Z\mid \parbox[t]{120mm}{$
\tn{dom}(t)$ is finite, $t$ is injective on its support, and
    \begin{itemize}
        \item $t(\{(i,i)\mid i\in\N^+\}) \subseteq \{0\}\cup\N^+$
        \item $t(\{(i,j)\mid i,j\in\N^+, i\ne j\})\subseteq \{0\}\cup \{-p\mid p\in\Podd\}$\Big\}.
    \end{itemize}
     
}$$

A partial oracle $w\in\Sigma^*$ is called $t$-valid for $t\in\calT$
if it satisfies the following properties.
\begin{itemize}
    \item[V1]   For all $i\in\N^+$ and all $x,y\in\N$, if $c(i,x,y)\in w$,
                then $F_i^w(x) = y$ and $y\in \oli{K^w}$.\\
                (meaning: if the oracle contains the codeword $c(i,x,y)$,
                then $F_i^w(x)$ outputs $y$ and $y \in \oli{K^w}$;
                hence, $c(i,x,y) \in w$ is a proof for $y \in \oli{K^w}$)
	\item[V2]   For all distinct $i,j\in\N^+$, if $t(i,j) = 0$, then there exists $x$ 
                such that (i) $M_i^w(x)$ and $M_j^w(x)$ definitely accept or
                (ii) $M_i^w(x)$ and $M_j^w(x)$ definitely reject.\\
                (meaning: for every extension of the oracle, $(M_i,M_j)$ 
                is not a $\cNP\cap\ccoNP$-machine.)
    \item[V3]   For all distinct $i,j\in\N^+$ with $t(i,j) = -p$ for some $p\in\Podd$ 
                and each $k\in\N^+$, it holds (i) $|\az^{p^k}\cap w|\le 1$ and
                (ii) if $w$ is defined for all words of length $p^k$, then
                $|\az^{p^k}\cap w| = 1$.
                \\(meaning: if $t(i,j) =-p$, then ensure that 
                $A_p =\oli{B_p}$ (i.e., $A_p\in\cNP\cap\ccoNP$) relative to the final oracle.)
    \item[V4]   For all $i\in\N^+$ with $t(i,i) = 0$, there exists $x$ such that $F_i^w(x)$ is
                defined and $F_i^w(x)\in K^v$ for all $v\sqsupseteq w$.\\
                (meaning: for every extension of the oracle, $F_i$ is not a proof system for $\oli{K}$)
    \item[V5]   For all $i\in\N^+$ and $x\in\N$ with $0 < t(i,i)\le c(i,x,F_i^w(x)) \!<\! |w|$,
                it holds $c(i,x,F_i^w(x)) \in w$.\\
                (meaning: if $t(i) > 0$, then from $t(i)$ on, we encode
                 $F_i$ into the oracle.\\
                Note that V5 is not in contradiction with V3 as $|c(\cdot,\cdot,\cdot)|$ is even.)
\end{itemize}

The subsequent claim follows directly from the definition of $t$-valid.
\begin{claim}\label{claim_90124780576}
Let $t,t'\in\calT$ such that $t'$ is an extension of $t$. 
For all oracles $w\in\az^*$, if $w$ is $t'$-valid, then $w$ is $t$-valid.
\end{claim}
\begin{claim}\label{claim_sandwich}
Let $t\in\calT$ and $u,v,w\in\az^*$ be oracles such that $u\sqsubseteq v\sqsubseteq w$
and both $u$ and $w$ are $t$-valid. Then $v$ is $t$-valid.
\end{claim}
\begin{proof}
$v$ satisfies V2 and V4 since $u$ satisfies these conditions. Moreover, $v$ satisfies
V3 as $w$ satisfies these conditions. 

Let $i\in\N^+$ and $x,y\in\N$ such that
$c(i,x,y)\in v$. Then $c(i,x,y)\in w$ and as $w$ is $t$-valid, we obtain by V1 that
$F_i^w(x) = y \in \oli{K^w}$. Claim~\ref{claim_7378194873} yields that $F_i^v(x)$ is defined
and $F_i^v(x) \in K^v\Leftrightarrow F_i^v(x)\in K^w$. This yields
that $F_i^v(x) = F_i^w(x) = y$ and $K^v(y) = K^w(y) = 0$. Thus, $v$ satisfies V1.

Now let $i\in\N^+$ and $x\in\N$ such that $0<t(i,i)\le c(i,x,F_i^v(x))<|v|$. Again,
by Claim~\ref{claim_7378194873}, $F_i^v(x)$ is defined and thus, $F_i^v(x) = F_i^w(x)$.
As $|v| \le |w|$ and $w$ is $t$-valid, we obtain by V5 that $c(i,x,F_i^v(x)) =
c(i,x,F_i^w(x))\in w$. Since $v\sqsubseteq w$ and $|v|>c(i,x,F_i^v(x))$, we obtain
$c(i,x,F_i^v(x))\in v$, which shows that $v$ satisfies V5.
\end{proof}

{\em Oracle construction.} Let $T$ be an enumeration of $(\N^+)^2\cup \{(i,j,r)\mid i\ne j, i,j,r\in\N^+\}$ 
having the property that $(i,j)$ appears
earlier than $(i,j,r)$ for all $i,j,r\in\N^+$ with $i\ne j$ (more formally, $T$ could be defined
as a function $\N \to (\N^+)^2\cup \{(i,j,r)\mid i\ne j, i,j,r\in\N^+\}$).
Each element of $T$
stands for a task. We treat the tasks in the order specified by $T$ and after
treating a task we remove it and possibly other tasks from $T$. We start
with the nowhere defined function $t_0$ and the $t_0$-valid oracle $w_0 = 
\varepsilon$. Then we define functions $t_1,t_2,\dots$ in $\calT$ such that $t_{i+1}$
is an extension of $t_i$ and partial oracles $w_0\sqsubsetneq w_1\sqsubsetneq w_2
\sqsubsetneq\dots$ such that each $w_i$ is $t_i$-valid. Finally, we choose
$O = \bigcup_{i = 0}^\infty w_i$ (note that $O$ is totally defined since in
each step we strictly extend the oracle).
We describe step $s>0$, which starts with some $t_{s-1}\in\calT$ and a $t_{s-1}$-valid oracle $w_{s-1}$
and chooses an extension $t_s\in\calT$ of $t_{s-1}$ and a $t_s$-valid $w_s\sqsupsetneq w_{s-1}$ (it will be argued later that
all these steps are indeed possible). 
Let us recall that each task is immediately deleted from $T$ after it is treated.

\begin{itemize}
    \item task $(i,i)$: Let $t' = t_{s-1}\cup \{(i,i)\mapsto 0\}$. If there
        exists a $t'$-valid $v\sqsupsetneq w_{s-1}$, 
        then let $t_s = t'$ and $w_s$ be the least $t'$-valid, partial oracle $\sqsupsetneq w_{s-1}$.
        Otherwise, let $t_s = t_{s-1}\cup \{(i,i) \mapsto |w_{s-1}|\}$ 
        and choose $w_s = w_{s-1}b$ with $b\in\{0,1\}$ 
        such that $w_s$ is $t_s$-valid.
        \\(meaning: try to ensure that $F_i$ is not a proof system for $\oli{K}$. If this is impossible,
        require that from now on the values of $F_i$ are encoded into the oracle.)
	\item task $(i,j)$ with $i\ne j$: Let $t' = 
        t_{s-1}\cup \{(i,j)\mapsto 0\}$. If there
        exists a $t'$-valid $v\sqsupsetneq w_{s-1}$, 
        then let $t_s = t'$, define $w_s$ to be the least $t'$-valid, partial oracle $\sqsupsetneq w_{s-1}$, 
        and delete all tasks $(i,j,\cdot)$ from $T$.
        Otherwise, let $z = |w_{s-1}|$, choose some $p\in\Podd$ 
        greater than $|z|$ with $-p\notin\ran(t_{s-1})$, let 
        $t_s = t_{s-1}\cup \{(i,j)\mapsto -p\}$,
        and choose $w_s = w_{s-1}b$ with $b\in\{0,1\}$
        such that $w_s$ is $t_s$-valid.
        \\(meaning: try to ensure that $(M_i,M_j)$ is not an $\cNP\cap\ccoNP$-machine. If this is
        impossible, then $L(M_i)$ inherently is in $\cNP\cap\ccoNP$ and we choose a sufficiently large prime $p$. 
        It will be made sure in the further construction that
        $A_p = \oli{B_p}$ and $A_p$ cannot be reduced to $L(M_i)$.)
    \item task $(i,j,r)$ with $i\ne j$: It holds $t_{s-1}(i,j) = -p$ for a prime $p\in\Podd$,
        since otherwise, this task would have been deleted in the treatment
        of task $(i,j)$. 
        Define $t_s = t_{s-1}$ and 
        choose a $t_s$-valid $w_s\sqsupsetneq w_{s-1}$
        such that for some $n\in\N^+$
        one of the following two statements holds:
        \begin{itemize}
            \item   $0^n\in A_p^v$ for all $v\sqsupseteq w_s$ 
                    and $M_i^{w_s}(F_r^{w_s}(0^n))$ definitely rejects.
            \item   $0^n\in B_p^v$ for all $v\sqsupseteq w_s$ 
                    and $M_j^{w_s}(F_r^{w_s}(0^n))$ definitely rejects.
        \end{itemize}
        (meaning: due to V3 it will hold $A_p = \oli{B_p}$ relative to the final oracle. By construction,
        relative to the final oracle it will hold $L(M_i) = \oli{L(M_j)}$. Hence, 
        the treatment of the task $(i,j,r)$ makes sure that it does not hold 
        $A_p\redm L(M_i)$ via $F_r$ relative to the final oracle.)
\end{itemize}
Observe that $t_s$ is always chosen in a way such that it is in $\calT$.
We now show that the construction is possible.
For that purpose, we first describe how a valid oracle can be extended
by one bit such that it remains valid.
\begin{claim}\label{claim_oracleextension}
Let $s\in\N$ and $w\in\az^*$ be a $t_s$-valid oracle with $w\sqsupseteq w_s$.
It holds for $z = |w|$:
    \begin{enumerate}
        \item\label{st_197400192} If $|z|$ is odd and for all $p\in\Podd$ 
        and $k\in\N^+$ with $-p\in\ran(t_s)$ it holds $|z|
        \ne p^k$, then $w0$ and $w1$ are $t_s$-valid.
        \item\label{st_10932471904} If there exist $p\in\Podd$ 
        and $k\in\N^+$ with $-p\in\ran(t_s)$
        such that $|z| = p^k$, $z\ne 1^{p^k}$, and 
        $w\cap \az^{p^k}=\emptyset$, then $w0$ and $w1$ are $t_s$-valid.
        \item\label{st_107924} If there exist $p\in\Podd$ and $k\in\N^+$ with $-p\in\ran(t_s)$
        such that $z = 1^{p^k}$ and $w\cap\az^{p^k} = \emptyset$, then $w1$ is $t_s$-valid.
        \item\label{st_178040812704} If $z = c(i,x,F_i^w(x))$ for $i\in\N^+$ and $x\in\N$ and $0<t_s(i,i)
        \le z$, then $w1$ is $t_s$-valid.
        \item\label{st_205832305} If $z = c(i,x,F_i^w(x))$ for $i\in\N^+$ and $x\in\N$, 
        at least one of the three
        conditions (i) $t_s(i,i)$ undefined, (ii) $t_s(i,i) = 0$, and (iii) $t_s(i,i) > z$ holds,
        and $F_i^w(x)\in \oli{K^w}$, then $w0$ and $w1$ are $t_s$-valid.
        \item\label{st_147890712034} In all other cases (i.e., none of the assumptions 
        in \ref{st_197400192}--\ref{st_205832305} holds) $w0$ is $t_s$-valid.
    \end{enumerate}
\end{claim}
\begin{proof}
First note that V2 and V4 are not affected by extending the oracle. So we only need
to consider V1, V3, and V5 in the following.

Let us show the following assertions.
\begin{align}
\parbox[c]{145mm}{$w0$ satisfies V1.}\label{eq_190472}\\
\parbox[c]{145mm}{If 
    (i) $z = c(i,x,F_i^w(x))$ for $i\in\N^+$ and $x\in\N$ with $F_i^w(x)\in \oli{K^w}$ or 
    (ii) $z$ has odd length,
    then $w1$ satisfies V1.}\label{eq_19023472}\\
\parbox[c]{145mm}{$w0$ satisfies V5 unless there exist $i\in\N^+$ and $x,y\in\N$ 
    such that
(i) $z = c(i,x,y)$,
(ii) $0<t_s(i,i) \le z$, and (iii) $F_i^w(x) = y$.}\label{eq_1904234342172}\\
\parbox[c]{145mm}{$w1$ satisfies V5.}\label{eq_342566190472}
\end{align}

(\ref{eq_190472}) and (\ref{eq_19023472}): Let $i'\in\N^+$ and $x',y'\in\N$ such that
$c(i',x',y')\in w$. Then, as $w$ is $t_s$-valid, by V1, $F_{i'}^w(x') = y'\in \oli{K^w}$ and by 
Claim~\ref{claim_7378194873}, $F_{i'}^{w}(x')$ is defined and $y'\in \oli{K^v}$ for all
$v\sqsupseteq w$. Hence, in particular, $F_{i'}^{wb}(x') = y'\in \oli{K^{wb}}$ for all $b\in
\{0,1\}$. This shows (\ref{eq_190472}). For the proof of (\ref{eq_19023472}) it remains 
to consider $z$. In case (ii) $w1$ satisfies V1 as $|z|$ is odd and each $c(i,x,y)$ has even length.
Consider case (i), i.e.,
$z = c(i,x,F_i^w(x))$ for $i\in\N^+$ and $x\in\N$ with $F_i^w(x)\in \oli{K^w}$.
Then by Claim~\ref{claim_7378194873}, $F_i^{w1}(x) =y\in \oli{K^{w1}}$, which shows that $w1$ satisfies
V1. This proves (\ref{eq_19023472}).

(\ref{eq_1904234342172}) and (\ref{eq_342566190472}): Let $i'\in\N^+$ and $x'\in\N$
such that $0<t_s(i',i')\le c(i',x',F_{i'}^w(x'))<|w|$. Then by Claim~\ref{claim_7378194873},
$F_{i'}^w(x')$ is defined and thus, $F_{i'}^{wb}(x') = F_{i'}^w(x')$ for all $b\in\{0,1\}$. 
As $w$ is $t_s$-valid, it holds $c(i',x',F_{i'}^w(x'))\in w$ and hence,
$c(i',x',F_{i'}^{wb}(x'))\in w \subseteq wb$ for all $b\in\{0,1\}$.
This shows (\ref{eq_342566190472}). In order to finish the proof of (\ref{eq_1904234342172}),
it remains to consider $z$. Assume $z = c(i,x,y)$ for some $i,x,y\in\N$ with $i>0$ and
$0<t_s(i,i)\le z$ (otherwise, $w0$ clearly satisfies V5).
If (iii) is wrong, then $F_i^w(x)\ne y$. By Claim~\ref{claim_7378194873}, this
computation is defined and hence, $F_i^{w0}(x)\ne y$, which is why $w0$ satisfies V5.
This shows (\ref{eq_1904234342172}).

We now prove the statements \ref{st_197400192}--\ref{st_147890712034}.
\begin{enumerate}
	\item Clearly $w0$ and $w1$ satisfy V3. Moreover, by (\ref{eq_190472}) and (\ref{eq_1904234342172}), the
    oracle $w0$ satisfies V1 and V5 (recall that the length of each $c(\cdot,\cdot,\cdot)$ is even). By
    (\ref{eq_19023472}) and (\ref{eq_342566190472}), the oracle $w1$ satisfies V1 and V5.
	\item By (\ref{eq_190472}), (\ref{eq_19023472}), (\ref{eq_1904234342172}), and (\ref{eq_342566190472}),
    the oracles $w0$ and $w1$ satisfy V1 and V5. As $z\ne 1^{p^k}$ and $w$ satisfies V3, the oracle $w0$ satisfies
    V3. As $w\cap\az^{p^k} = \emptyset$, the oracle $w1$ satisfies V3.
	\item By (\ref{eq_19023472}) and (\ref{eq_342566190472}),
    the oracle $w1$ satisfies V1 and V5. As $w\cap\az^{p^k} = \emptyset$, the oracle $w1$ satisfies V3.
	\item As $|z|$ is even, $w1$ satisfies V3. By (\ref{eq_342566190472}), $w1$ satisfies V5. It remains to
    argue that $w1$ satisfies V1. In order to apply (\ref{eq_19023472}), which will immediately show that
    $w1$ satisfies V1, it is sufficient to prove $y:=F_i^w(x)\in\oli{K^w}$. For a contradiction assume $y\in K^w$.
    Let $s'$ be the step that treats the task $(i,i)$. Note $s' < s$ since $t_s(i,i)$ is defined. By
    Claim~\ref{claim_90124780576}, $w$ is $t_{s'-1}$-valid. As by Claim~\ref{claim_7378194873} the computation
    $F_i^w(x)$ is defined and $y\in K^v$ for all $v\sqsupseteq w$, the oracle $w$ is even $t$-valid 
    for $t = t_{s'-1} \cup \{(i,i)\mapsto 0\}$. But then the construction would have chosen $t_{s'} = t$,
    in contradiction to $t_s(i,i) > 0$.
	\item As $|z|$ is even, $w0$ and $w1$ satisfy V3. By (\ref{eq_190472}), (\ref{eq_19023472}),
    and (\ref{eq_342566190472}), $w0$ satisfies V1 and $w1$ satisfies both V1 and V5.
    Moreover, (\ref{eq_1904234342172}) can be applied since each of the conditions (i)--(iii)
    of statement \ref{st_205832305} implies that
    condition (ii) of (\ref{eq_1904234342172}) does not hold.
    Thus, $w0$ satisfies V5.
	\item By (\ref{eq_190472}), $w0$ satisfies V1. If $w0$ does not satisfy V3, then there exist
    $p\in\Podd$ with $-p\in\tn{ran}(t_s)$ and $k>0$ such that $w\cap\az^{p^k} = \emptyset$ and $z = 1^{p^k}$,
    but this case is covered by statement \ref{st_107924} of the current claim. If $w0$ does not satisfy V5, then
    by (\ref{eq_1904234342172}), there exist $i\in\N^+$ and $x,y\in\N$ such that (i) $z = c(i,x,y)$,
    (ii) $0<t_s(i,i) \le z$, and (iii) $F_i^w(x) = y$. This case, however, is covered by
    statement \ref{st_178040812704} of the current claim.
\end{enumerate}

This finishes the proof of Claim~\ref{claim_oracleextension}.
\end{proof}

In order to show that the above construction is possible, assume that 
it is not possible and let $s>0$
be the least number, where it fails.

If step~$s$ treats a task $t\in(\N^+)^2$, then $t_{s-1}(t)$ is
not defined, since the value of $t$ is defined in the 
unique treatment of the task $t$. Hence, $t'$ is well-defined. If $t_s(t)$ is chosen 
to be 0, then the construction clearly is possible.
Otherwise, due to the choice of $t_s(t)$, the 
$t_{s-1}$-valid oracle $w_{s-1}$ is
even $t_s$-valid and Claim~\ref{claim_oracleextension} 
ensures that there exists a $t_s$-valid
$w_{s-1}b$ for some $b\in\{0,1\}$. Hence, the construction does not fail in step~$s$, 
a contradiction.

For the remainder of the proof that the construction above is possible
we assume that step~$s$ treats a task 
$(i,j,r)\in\{(i,j,r)\mid i\ne j, i,j,r\in\N^+\}$.

Then $t_s = t_{s-1}$ and $t_s(i,j)
= -p$ for some $p\in\Podd$. Let $\gamma$ be the polynomial given by $x \mapsto (x^r+r)^{i+j} + i+j$
and choose $k\in\N^+$
such that for $n = p^k$
\begin{equation}\label{eq_01847012}
2^{n-1} > 2\cdot \gamma(n)
\end{equation}
and $w_{s-1}$ is not defined for any words of length $n$.
Note that $\gamma(n)$ is greater than the running time of each
of the computations $M_i^D(F_r^D(0^n))$ and $M_j^D(F_r^D(0^n))$
for each oracle $D$.

We define $u\sqsupseteq w_{s-1}$ to be the minimal $t_s$-valid oracle 
that is defined for all words of length $<n$. Such an oracle exists by 
Claim~\ref{claim_oracleextension}. 

Moreover, for $z\in \az^n$, let $u_z\sqsupsetneq u$
be the minimal $t_s$-valid oracle with $u_z\cap\az^n = \{z\}$ that
is defined for all words of length $\le \gamma(n)$. 
Such an oracle exists by 
Claim~\ref{claim_oracleextension}: 
first, starting from $u$ we extend the current oracle bitwise such that
(i) it remains $t_s$-valid, (ii) it is defined for precisely the words of length $\le n$,
and (iii) its intersection with $\az^n$ equals $\{z\}$. This is possible
by \ref{st_10932471904}, \ref{st_107924}, and \ref{st_147890712034} of Claim~\ref{claim_oracleextension}. Then by Claim~\ref{claim_oracleextension},
the current oracle can be extended bitwise without losing its $t_s$-validity until it is
defined for all words of length $\le\gamma(n)$.

We define a further oracle $v$ that will be crucial in the following. 
Let $s'$ be the step that treats the task $(i,j)$. As $t_s(i,j)$ is defined,
it holds $s' < s$. By Claim~\ref{claim_90124780576}, the oracle $u$ is $t_{s'-1}$-valid.
In order to define $v$, we need the following two properties (\ref{eq_0135973190}) and
(\ref{eq_2509727590}) that we also need in different
contexts and therefore, define in a general way. Let $w\sqsupseteq u$ be a $t_{s'-1}$-valid
oracle. We say that $w$ satisfies  property (\ref{eq_0135973190}) if
\begin{align}\label{eq_0135973190}
\parbox[c]{145mm}{
for all $i',x\in\N$ with $i'>0$, 
$t_s(i',i')>0$ and $|u| \le c(i',x,F_{i'}^{w}(x)) < |w|$, if $F_{i'}^{w}(x)\in \oli{K^w}$, then
$c(i',x,F_{i'}^{w}(x))\in w$.
}
\end{align}
Moreover, $w$ satisfies  property (\ref{eq_2509727590}) if
\begin{align}\label{eq_2509727590}
\parbox[c]{145mm}{
for all $p'\in\Podd$ with $-p\in\ran(t_s)$, 
\begin{enumerate}[noitemsep, topsep=0pt]
	\item[(i)] $w\cap \az^{{p'}^\kappa} \subseteq \{1^{{p'}^\kappa}\}$ for all $\kappa > 0$ with
    $n < {p'}^\kappa$ and
    \item[(ii)] $w\cap \az^{{p'}^\kappa} = \{1^{{p'}^\kappa}\}$ for all $\kappa > 0$ 
    with $n < {p'}^\kappa$ and $1^{{p'}^\kappa} < |w|$.
\end{enumerate}
}
\end{align}

Now we define $v\sqsupsetneq u$ to be the minimal $t_{s'-1}$-valid oracle that is defined
for all words of length $\le\gamma(n)$ and satisfies properties (\ref{eq_0135973190}) and
(\ref{eq_2509727590}).
Let us argue that such an oracle exists. Clearly $u$ satisfies properties (\ref{eq_0135973190}) and
(\ref{eq_2509727590}). The second statement of the following claim shows that $v$ is well-defined.
\begin{claim}\label{claim_41207012420}
\begin{enumerate}
    \item\label{st_104871022422} For all $t_{s'-1}$-valid oracles $w$ and $w'$ with $u\sqsubseteq w\sqsubseteq w'$,
    if $w$ satisfies property~(\ref{eq_0135973190}) and $w'$ does not satisfy property~(\ref{eq_0135973190}),
        then there exists $|w|\le \alpha < |w'| $ such that
        $\alpha = c(i',x,F_{i'}^{w'}(x))$ for $i',x\in\N$
        with $i'>0$ and $t_s(i',i') > 0$, $F_{i'}^{w'}(x)\in\oli{K^{w'}}$, and $\alpha\notin w'$.
    \item\label{st_12470} For each $t_{s'-1}$-valid oracle $w\sqsupseteq u$ that satisfies 
    properties (\ref{eq_0135973190}) and
(\ref{eq_2509727590}) there exists $b\in\{0,1\}$ such that $wb$ is $t_{s'-1}$-valid and satisfies properties
(\ref{eq_0135973190}) and (\ref{eq_2509727590}).
    \item\label{st_0194270} Let $w\sqsupseteq u$ be $t_{s'-1}$-valid. 
        If $w$ satisfies properties (\ref{eq_0135973190}) and (\ref{eq_2509727590}), then
        each $w'$ with $u\sqsubseteq w'\sqsubseteq w$ satisfies properties 
        (\ref{eq_0135973190}) and (\ref{eq_2509727590}).
\end{enumerate}
\end{claim}
\begin{proof}
\begin{enumerate}[wide, labelwidth=!, labelindent=0pt,topsep=0pt,partopsep=0pt]
\item Since $w'$ does not satisfy property~(\ref{eq_0135973190}), there exists $|u|\le
    \alpha = c(i',x,F_{i'}^{w'}(x)) < |w'|$
    for $i',x\in\N$ with $i'>0$ and $t_s(i',i')>0$ such that $F_{i'}^{w'}(x)\in \oli{K^{w'}}$ and $\alpha\notin w'$.
    For a contradiction we assume $\alpha < |w|$. Then Claim~\ref{claim_7378194873}
    yields $F_{i'}^w(x) = F_{i'}^{w'}(x) \in\oli{K^w}$. From $\alpha<|w|$, $w\sqsubseteq w'$, and $\alpha\notin w'$
    it follows $\alpha\notin w$, which contradicts the assumption that $w$ satisfies property~(\ref{eq_0135973190}).
\item We study several cases depending on $\alpha = |w|$ (i.e., $\alpha$ is the least word that
$w$ is not defined for).
\begin{itemize}
	\item If $\alpha$ is of the form $c(i',x,F_{i'}^w(x))$ for $i',x\in\N$ with $i'>0$ and $t_s(i',i')>0$
    such that $F_{i'}^w(x)\in\oli{K^w}$, then we choose $b = 1$. 
    The statements~\ref{st_178040812704} and \ref{st_205832305} of
    Claim~\ref{claim_oracleextension}
    state that the oracle $wb$ is $t_{s'-1}$-valid (recall
    that by construction $t_s(i',i') \le |u| \le |w| = \alpha$ and 
    note that we apply Claim~\ref{claim_oracleextension} for the parameter $s'-1$). 
    \item If $\alpha$ has length
    ${p'}^\kappa$ for some $p'\in\Podd$ with $-p'\in \tn{ran}(t_s)$ and $\kappa > 0$,
    then we choose $b = 1$ if $\alpha = 1^{{p'}^\kappa}$ and $b = 0$ otherwise.
    Since $w$ satisfies property~(\ref{eq_2509727590}), it holds $w\cap\az^{{p'}^\kappa} = \emptyset$. Hence,
    the statements~\ref{st_197400192}, \ref{st_10932471904}, and \ref{st_107924} of 
    Claim~\ref{claim_oracleextension} state that the oracle $wb$ is  $t_{s'-1}$-valid
    (again, note that we apply Claim~\ref{claim_oracleextension} for the parameter $s'-1$).
    \item In all other cases Claim~\ref{claim_oracleextension} guarantees that we can choose $b\in\{0,1\}$
    such that $wb$ is $t_{s'-1}$-valid. 
\end{itemize}
By the choice of $b$, the oracle $wb$ satisfies property (\ref{eq_2509727590}). If $wb$ does not
satisfy property~(\ref{eq_0135973190}), then by
statement~\ref{st_104871022422} of the current claim, $\alpha = c(i',x,F_{i'}^{wb}(x))$ for $i',x\in\N$
with $i'>0$ and $t_s(i',i') > 0$, $F_{i'}^{wb}(x)\in\oli{K^{wb}}$, and $\alpha\notin wb$. 
Claim~\ref{claim_7378194873}, however, yields that then even 
$F_{i'}^{w}(x) = F_{i'}^{wb}(x) \in\oli{K^{w}}$. But then we would have chosen $b = 1$ above, in
contradiction to $\alpha\notin wb$.
\item As $w'\sqsubseteq w$ and $w$ satisfies property~(\ref{eq_2509727590}), $w'$ satisfies
property~(\ref{eq_2509727590}).
We argue that $w'$ satisfies property~(\ref{eq_0135973190}).
Let $i',x\in\N$ with $i'> 0$ and $t_s(i',i')>0$
such that $|u| \le c(i',x,F_{i'}^{w'}(x)) < |w'|$ and $F_{i'}^{w'}(x)\in \oli{K^{w'}}$.
As $c(i',x,F_{i'}^{w'}(x)) < |w'|$, Claim~\ref{claim_7378194873} 
yields $F_{i'}^{w}(x) = F_{i'}^{w'}(x)\in \oli{K^w}$. As $w$ satisfies
property~(\ref{eq_0135973190}), it holds $c(i',x,F_{i'}^{w'}(x))\in w$. Since $w'\sqsubseteq w$ and
$c(i',x,F_{i'}^{w'}(x)) < |w'|$, we obtain $c(i',x,F_{i'}^{w'}(x)) \in w'$. Hence, $w'$ satisfies
property~(\ref{eq_0135973190}).
\end{enumerate}
This finishes the proof of Claim~\ref{claim_41207012420}.
\end{proof}

Note that by the choice of $v$, it holds $v\cap\az^n = \emptyset$ (cf.\
Claim~\ref{claim_oracleextension}.\ref{st_197400192} and recall that $t_{s'-1}$
is not defined for the pair $(i,j)$).

\begin{claim}\label{claim_1047120472314}
Let $w\in\{v\}\cup\{u_z\mid z\in\az^n\}$. 
\begin{enumerate}
	\item\label{st_172304231740} For each $\alpha\in w\cap \az^{>n}$ one of the following statements holds.
        \begin{itemize}
            \item   $\alpha = c(i',x,F_{i'}^{w}(x))$ for some $i'\in\N^+$ and $x\in\N$ with 
                    $0<t_s(i',i') \le c(i',x,F_{i'}^{w}(x))$ and $F_{i'}^{w}(x) \in \oli{K^{w}}$.
            \item   $\alpha = 1^{{p'}^\kappa}$ for some $p'\in\Podd$ with $-p'\in\ran(t_s)$ 
                    and some $\kappa>0$.
        \end{itemize}
    \item\label{st_14790120} For all $p'\in\Podd$ 
        with $-p'\in\ran(t_s)$ and all $\kappa>0$, 
        if $n< {p'}^\kappa \le \gamma(n)$, then
        $w \cap \az^{{p'}^\kappa} = \{1^{{p'}^\kappa}\}$.
    \item\label{st_10471023} For all $z\in \az^n$ and all $\alpha\in u_z - v$ it holds
    $\alpha = c(i',x,F_{i'}^{u_z}(x))$ for some $i'\in\N^+$ and $x\in\N$ with 
    $0<t_s(i',i') \le c(i',x,F_{i'}^{u_z}(x))$ and $F_{i'}^{u_z}(x) \in \oli{K^{u_z}}$.
    \item\label{st_105897102}For all $z\in \az^n$ and all $\alpha\in v - u_z$ it holds
    $\alpha = c(i',x,F_{i'}^{v}(x))$ for some $i'\in\N^+$ and $x\in\N$ with 
    $0<t_s(i',i') \le c(i',x,F_{i'}^{v}(x))$ and $F_{i'}^{v}(x)\in \oli{K^{v}}$.
\end{enumerate}
\end{claim}
\begin{proof}
\begin{enumerate}[wide, labelwidth=!, labelindent=0pt,topsep=0pt,partopsep=0pt]
\item We first argue for the case $w = u_z$ for some $z\in\az^n$.
Let $\alpha\in u_z\cap\az^{>n}$. Moreover, let $u'$ be the prefix of $u_z$ that
has length $\alpha$, i.e., $\alpha$ is the least word that $u'$ is not defined for.
In particular, it holds $u'\cap\az^{\le n} = u_z\cap\az^{\le n}$ and thus, 
$u'\cap\az^n = \{z\}$.
As $u\sqsubseteq u'\sqsubseteq u_z$ and both $u$ and $u_z$ are $t_s$-valid, 
Claim~\ref{claim_sandwich} yields that $u'$ is also $t_s$-valid.
\\\indent Let us apply Claim~\ref{claim_oracleextension} to the oracle
$u'$. If one of the cases \ref{st_197400192}, \ref{st_10932471904},
\ref{st_205832305}, and \ref{st_147890712034} can be applied, then $u'0$ is $t_s$-valid
and can be extended to a $t_s$-valid oracle $u''$ with $|u''| = |u_z|$ by 
Claim~\ref{claim_oracleextension}. As $u''$ and $u_z$ agree on all
words $<\alpha$ and $\alpha\in u_z-u''$, we obtain $u'' < u_z$ and due to $u'\sqsubseteq u''$
we know that $u''\cap\az^n = \{z\}$. This is 
a contradiction to the choice of $u_z$ (recall that $u_z$ is the 
minimal $t_s$-valid oracle that is defined for all words of length $\le \gamma(n)$ and 
that satisfies $u_z\cap\az^n = \{z\}$).
\\\indent Hence, none of the cases \ref{st_197400192}, \ref{st_10932471904},
\ref{st_205832305}, and \ref{st_147890712034} of 
Claim~\ref{claim_oracleextension} can be applied, i.e., either (i) Claim~\ref{claim_oracleextension}.\ref{st_107924}
or (ii) Claim~\ref{claim_oracleextension}.\ref{st_178040812704} can be applied.
Hence, either (i) $\alpha = 1^{{p'}^\kappa}$
for some $p'\in\Podd$ and $\kappa>0$ with $-p'\in\ran(t_s)$ or (ii) 
$\alpha = c(i',x,F_{i'}^{u_z}(x))$ for $i'\in\N^+$ and $x\in\N$ 
with $0<t_s(i',i') \le \alpha$. In the latter
case, as $\alpha\in u_z$ and $u_z$ is $t_s$-valid, we obtain from V1 that $F_{i'}^{u_z}(x) \in \oli{K^{u_z}}$.

\smallskip The arguments for the case $w = v$ are similar:
Let $\alpha\in v\cap\az^{>n}$. Moreover, let $v'$ be the prefix of $v$ that
has length $\alpha$, i.e., $\alpha$ is the least word that $v'$ is not defined for.
As $u\sqsubseteq v'\sqsubseteq v$ and both $u$ and $v$ are $t_{s'-1}$-valid, 
Claim~\ref{claim_sandwich} yields that $v'$ is also $t_{s'-1}$-valid.
Moreover, by Claim~\ref{claim_41207012420}.\ref{st_0194270}, $v'$ satisfies properties 
(\ref{eq_0135973190}) and (\ref{eq_2509727590}).
\\\indent Let us apply Claim~\ref{claim_oracleextension} to the oracle
$v'$ (with the parameter $s'-1$). If one of the cases \ref{st_197400192}, \ref{st_10932471904},
\ref{st_205832305}, and \ref{st_147890712034} can be applied, then $v'0$ is $t_{s'-1}$-valid. 
\\First, assume that it does not hold that $v'0$ satisfies properties
(\ref{eq_0135973190}) and (\ref{eq_2509727590}). 
If $wb$ does not satisfy property~(\ref{eq_2509727590}), then
$\alpha = 1^{{p'}^\kappa}$ for some 
$p'\in\Podd$ with $-p'\in\tn{ran}(t_s)$ and $\kappa>0$.
If $wb$ does not satisfy property~(\ref{eq_0135973190}), then it holds
by Claim~\ref{claim_41207012420}.\ref{st_104871022422} that
$\alpha = c(i',x,F_{i'}^{v'0}(x))$
for some $i',x\in\N$ with $i'>0$ and $t_s(i',i')>0$ such that
$F_{i'}^{v'0}(x)\in \oli{K^{v'0}}$.
By Claim~\ref{claim_7378194873} and by $|\alpha|\le |v'|$, we obtain
$F_{i'}^{v'}(x) = F_{i'}^{v'0}(x)\in \oli{K^{v'}}$. Moreover, by construction,
$\alpha > |u| \ge t_s(i',i')$. Hence, under the assumption that $v'0$ does not satisfy property~(\ref{eq_0135973190}) or $v'0$ does not satisfy property~(\ref{eq_2509727590}), we obtain that $\alpha$ is of the form
described by the current claim.
\\Now we consider the case that $v'0$ 
satisfies properties (\ref{eq_0135973190}) and (\ref{eq_2509727590}) and show that this assumption
leads to a contradiction.
By iteratively applying 
Claim~\ref{claim_41207012420}.\ref{st_12470} we extend $v'0$
to a $t_{s'-1}$-valid oracle $v''$ that satisfies $|v''| = |v|$ and properties 
(\ref{eq_0135973190}) and (\ref{eq_2509727590}). As $v''$ and $v$ agree on 
all words $<\alpha$ and $\alpha\in v - v''$,
it holds $v'' < v$, in contradiction to the choice of $v$ 
(recall that $v$ is the minimal $t_{s'-1}$-valid oracle
$\sqsupsetneq u$ that is defined for all words of length 
$\le\gamma(n)$ and satisfies properties (\ref{eq_0135973190}) 
and (\ref{eq_2509727590})).

In order to finish the proof of statement~\ref{st_172304231740}, it remains to consider the
cases that Claim~\ref{claim_oracleextension}.\ref{st_107924} or Claim~\ref{claim_oracleextension}.\ref{st_178040812704}
can be applied to $v'$. This means that either (i) there exist $p'\in\Podd$ and $\kappa\in\N^+$ with $-p\in\ran(t_{s'-1})\subseteq \ran(t_s)$
such that $z = 1^{{p'}^\kappa}$, or (ii) $\alpha = c(i',x,y)$ for $i'\in\N^+$ and $x,y\in\N$ with $0<t_{s'-1}(i',i')
= t_s(i',i')\le \alpha$. In the latter
case, as $\alpha\in v$ and $v$ is $t_{s'-1}$-valid, we obtain from V1 that $F_{i'}^{v}(x) = y\in \oli{K^{v}}$.

\item The statement is true in case $w = v$ as $v$ satisfies property (\ref{eq_2509727590}).
Let us argue for the case $w = u_z$ for some $z\in\az^n$.
As $-p'\in\ran(t_s)$, $u_{z}$ is $t_s$-valid, and $u_z$ is defined
for all words of length ${p'}^\kappa$, V3 yields that
there exists $\beta\in\az^{{p'}^\kappa}\cap u_{z}$. 
Let $\beta$ be the minimal element of $\az^{{p'}^\kappa}\cap u_{z}$.
It suffices to show $\beta = 1^{{p'}^\kappa}$.
For a contradiction, we assume $\beta < 1^{{p'}^\kappa}$.
Let $u'$ be the prefix of $u_z$ that is defined for exactly the words
$<\beta$. Then $u\sqsubseteq u'\sqsubseteq u_z$ and
both $u$ and $u_z$ are $t_s$-valid. Hence, by Claim~\ref{claim_sandwich},
the oracle $u'$ is $t_s$-valid as well.
\\\indent By Claim~\ref{claim_oracleextension}, $u'$ can be extended
to a $t_s$-valid oracle $u''$ that satisfies $|u''| = |u_z|$ 
and $u''\cap\az^{{p'}^\kappa} = \{ 1^{{p'}^\kappa}\}$.
Then $\beta\in u_{z}-u''$. As the oracles $u''$ and $u_{z}$ agree on all words $<\beta$,
we have $u'' < u_{z}$ and $u''\cap\az^n = \{z\}$, in contradiction
to the choice of $u_{z}$ (again, recall that $u_z$ is the 
minimal $t_s$-valid oracle that is defined for all words of length $\le \gamma(n)$ and 
that satisfies $u_z\cap\az^n = \{z\}$).

\item This statement follows from the statements \ref{st_172304231740} and \ref {st_14790120}.

\item This statement follows from the statements \ref{st_172304231740} and \ref {st_14790120}.
\end{enumerate}
This finishes the proof of Claim~\ref{claim_1047120472314}.
\end{proof}

Let us study the case that both computations
$M_i^{v}(F_r^{v}(0^n))$ and $M_j^{v}(F_r^{u_z}(0^n))$ reject.
Then they even definitely reject as $v$ is defined for all
words of length $\le\gamma(n)$. But then $v$ is not only $t_{s'-1}$-valid but even $t$-valid for
$t = t_{s'-1}\cup\{(i,j)\mapsto 0\}$ and then the construction would have chosen $t_{s'} = t$, in contradiction
to $t_s(i,j) = -p<0$. Hence one of the computations $M_i^{v}(F_r^{v}(0^n))$ and $M_j^{v}(F_r^{u_z}(0^n))$
accepts and thus, even definitely accepts. By symmetry, it suffices to consider the
case that $M_i^{v}(F_r^{v}(0^n))$ definitely accepts.

Let $U$ be the set of all those oracle queries of the least accepting path 
of $M_i^{v}(F_r^{v}(0^n))$ that are of length $\ge n$. Observe $\ell(U) \le \gamma(n)$.
Moreover, define $Q_0(U) = U$ and for $m\in\N$,
\begin{align*}
Q_{m+1}(U) = &\bigcup_{\substack{c(i',x,y)\in Q_m(U)\\
i',x,y\in\N, i'>0}} \Big[\{q\in \az^{\ge n}\mid 
\text{$q$ is queried by $F_{i'}^{v}(x)$}\}\cup\\
&\{q\in \az^{\ge n}\mid 
\parbox[t]{110mm}{$y = \pairing{0^{i''}, 0^{|x'|^{i''}+i''}, x'}$ for some $i''>0$ and $x'\in\N$, 
$M_{i''}^{v}(x')$ has an accepting path, and $q$ is queried by the least such path$\}\Big]$.} 
\end{align*}

Let $Q(U) = \bigcup_{m \in\N} Q_m(U)$. Note that all words in
$Q(U)$ have length $\ge n$.
\begin{claim}\label{claim_2140771289032}
$\ell(Q(U)) \le 2 \ell(U) \le 2\gamma(n)$ and
the length of each word in $Q(U)$ is $\le \gamma(n)$.
\end{claim}
\begin{proof}
We show that for all $m\in\N$,
$\ell(Q_{m+1}(U))\le \nicefrac{1}{2}\cdot \ell(Q_m(U))$. Then
$\sum_{m = 0}^s \nicefrac{1}{2^m} \le 2$ for all $s\in\N$
implies $\ell(Q(U))\le 2\cdot\ell(U)\le 2\gamma(n)$. Moreover, from 
$\ell(U)\le \gamma(n)$ and $\ell(Q_{m+1}(U))\le \nicefrac{1}{2}\cdot \ell(Q_m(U))$
the second part of the claim follows.

Let $m\in \N$ and consider an arbitrary element $\alpha$ of $Q_m(U)$. If $\alpha$
is not of the form $c(i',x,y)$ for $i'\in\N^+$ and $x,y\in\N$,
then $\alpha$ generates no elements in $Q_{m+1}(U)$.
Assume $\alpha = c(i',x,y)$ for $i'\in\N^+$ and $x,y\in \N$ with
$y=\pairing{0^{i''}, 0^{|x'|^{i''}+i''}, x'}$ for $i''\in\N^+$ and $x'\in\N$.
The computation $F_{i'}^v(x)$ runs for at most 
$|x|^{i'}+i' < \nicefrac{|\alpha|}{4}$ steps, where \eqq{$<$} holds
by (\ref{eq_140712041003}). 
Hence, the set of queries $Q$ of $F_{i'}^{v}(x)$ satisfies
$\ell(Q)\le\nicefrac{|\alpha|}{4}$. 

Moreover, the computation $M_{i''}^v(x)$ runs for less than
$|y| < \nicefrac{|\alpha|}{4}$ steps, where again ``$<$'' holds by (\ref{eq_140712041003}). 
Hence, for the set $Q$ of queries
of the least accepting path of the computation $M_{i''}^{v}(x)$ (if such a path exists) we have
$\ell(Q) \le \nicefrac{|\alpha|}{4}$.

Consequently,
\begin{align*}\ell(Q_{m+1}(U))& \le  \sum_{\substack{c(i',x,y)\in Q_m(U)\\
i',x,y\in\N, i'>0}}
\Big[\underbrace{\ell\big(\{q\in \az^{\ge n}\mid 
\text{$q$ is queried by $F_{i'}^{v}(x)$}\}\big)}_{
\le \nicefrac{|c(i',x,y)|}{4}} +
\\&
\phantom{\le \sum_{\substack{c(i',x,y)\in Q_m(U)\\
i',x,y\in\N, i'>0}}}  \underbrace{\ell\big(\{q\in \az^{\ge n}\mid 
\parbox[t]{85mm}{$y = \pairing{0^{i''}, 0^{|x'|^{i''}+i''}, x'}$ for some $i''>0$ and $x'\in\N$, 
$M_{i''}^{v}(x')$ has an accepting path, and $q$ is queried by the least such path$\}\big)\Big]$}}_{
\le \nicefrac{|c(i',x,y)|}{4}}\\
&\le \sum_{\substack{c(i',x,y)\in Q_m(U)\\
i',x,y\in\N, i'>0}} \nicefrac{|c(i',x,y)|}{2}
\\&\le \nicefrac{\ell(Q_m(U))}{2},
\end{align*}
which finishes the proof of Claim~\ref{claim_2140771289032}.
\end{proof}

For $z$ even we say that $u_z$ and $v$ conflict if there exists
$\alpha\in Q(U)$ with $\alpha\in u_z\triangle v$. In that case we say
that $u_z$ and $v$ conflict in $\alpha$. As $u_z\sqsupseteq u$ and $v\sqsupseteq u$,
either $u_z$ and $v$ conflict in a word of length $\ge n$, or they do not conflict at all.

\begin{claim}\label{claim_1807516402141}
There exists an even $z\in\az^n$ such that $u_z$ and $v$ do not conflict.
\end{claim}
\begin{proof}
Let $z\in\az^n$ be even such that $u_z$ and $v$ conflict. We show that then
$u_z$ and $v$ conflict in $z$.
Let $\alpha\in Q(U)$ be the
least word of length $>n$ that $u_z$ and $v$ conflict in. Then $\alpha\in v\triangle u_z$. We study two cases.
\begin{itemize}
	\item Assume $\alpha\in u_z - v$. By Claim~\ref{claim_1047120472314}.\ref{st_10471023},
    it holds $\alpha = c(i',x,y)$ for some $i'\in\N^+$ and $x,y\in\N$ with 
    $0<t_s(i',i') \le c(i',x,F_{i'}^{u_z}(x))$ and $F_{i'}^{u_z}(x) = y \in \oli{K^{u_z}}$.
    
    First assume $F_{i'}^v(x) \ne y$. Then there is one query $q$ of $F_{i'}^v(x)$
    that is in $v\triangle u_z$ (otherwise, $F_{i'}^{u_z}(x)$ and $F_{i'}^v(x)$ would output
    the same value). As $v$ and $u_z$ agree on all words of length $< n$, it holds $|q|\ge n$. 
    Hence, by $\alpha\in Q(U)$ and the definition of $Q(U)$, it holds $q\in Q(U)$. 
    As $|q|\le |x|^{i'}+i' < |c(i',x,y)| = |\alpha|$
    and $\alpha$ is the least word of length $>n$ in $Q(U)$ that $v$ and $u_z$ conflict in, 
    it holds $|q| = n$. Hence, $v$ and $u_z$ conflict in a word of length $n$.
    
    Now assume $F_{i'}^v(x) = y$. As $\alpha\notin v$ and $v$ satisfies property~(\ref{eq_0135973190}), it holds
    $y\notin \oli{K^v}$. As $y\in K^v$, $y$ is of the form $\pairing{0^{i''},
    0^{|x'|^{i''}},x'}$ for some $i''> 0$ and $x'\in\N$. From $y\in K^v$ it follows that
    the computation $M_{i''}^v(x')$ has an accepting path and all queries $q$ of length $\ge n$
    that are asked on the least such path are in $Q(U)$. However, $y\in \oli{K^{u_z}}$ yields that
    there is some query $q$ on the least accepting path of $M_{i''}^v(x')$ that is in $v\triangle u$
    (otherwise, $M_{i''}^{u_z}(x')$ would accept as well). As $v$ and $u_z$ agree on all words of length $< n$,
    it holds $|q|\ge n$ and by this, $\alpha\in Q(U)$, and the definition of $Q(U)$, it holds $q\in Q(U)$.
    Since $|q|\le |x|^{i''}+i'' < |y| < |c(i',x,y)| = |\alpha|$ 
    and $\alpha$ is the least word of length $>n$ in $Q(U)$ that $v$ and $u_z$ conflict in, it holds $|q| = n$. Hence,
    $v$ and $u_z$ conflict in a word of length $n$.
	\item Assume $\alpha\in v - u_z$. By Claim~\ref{claim_1047120472314}.\ref{st_105897102},
    it holds $\alpha = c(i',x,F_{i'}^{v}(x))$ for some $i'\in\N^+$ and $x\in\N$ with 
    $0<t_s(i',i') \le c(i',x,F_{i'}^{v}(x))$ and $F_{i'}^{v}(x)\in \oli{K^{v}}$.
    If $F_{i'}^{u_z}(x) = F_{i'}^{v}(x)$, then
    by V5, we have $\alpha\in u_z$,
    a contradiction. Hence, $F_{i'}^{u_z}(x) \ne F_{i'}^{v}(x)$. Then there is one query $q$ of $F_{i'}^v(x)$
    that is in $v\triangle u_z$ (otherwise, $F_{i'}^{u_z}(x)$ and $F_{i'}^v(x)$ would output
    the same value). As $v$ and $u_z$ agree on all words of length $< n$, it holds $|q|\ge n$. By this,
    $\alpha\in Q(U)$, and the definition of $Q(U)$, it holds $q\in Q(U)$. 
    Since $|q|\le |x|^{i'}+i' < |c(i',x,F_{i'}^{v}(x))| = |\alpha|$
    and $\alpha$ is the least
    word of length $>n$ in $Q(U)$ that $v$ and $u_z$ conflict in, it holds $|q| = n$. Hence,
    $v$ and $u_z$ conflict in a word of length $n$.
\end{itemize}
In both cases $v$ and $u_z$ conflict in a word of length $n$. As $v\cap\az^n = \emptyset$
and $u_z\cap\az^n = \{z\}$, the oracles $v$ and $u_z$ conflict in $z$ and in particular,
$z\in Q(U)$.

From $|Q(U)|\le\ell(Q(U))\le 2\gamma(n)$ (cf.\ Claim~\ref{claim_2140771289032}) we obtain
that there are at most $2\gamma(n)$ even words $z\in\az^n$ that $v$ and $u_z$ conflict in.
As by (\ref{eq_01847012}), it holds $|\{z\in\az^n\mid \text{$z$ even}\}| = 2^{n-1} > 2\gamma(n)$,
the proof of Claim~\ref{claim_1807516402141} is complete.
\end{proof}
As guaranteed by Claim~\ref{claim_1807516402141}, we can now choose some even $z\in\az^n$ such that
$v$ and $u_z$ do not conflict. As all queries of the least accepting path of
$M_i^v(F_r^v(0^n))$ are in $U\subseteq Q(U)$ and
$v$ and $u_z$ agree on all these queries, the computation $M_i^{u_z}(F_r^{u_z}(0^n))$ accepts. Since
$u_z$ is defined for all words of length $\le \gamma(n)$, the computation even definitely accepts.
Note that the computation $M_j^{u_z}(F_r^{u_z}(0^n))$ is defined as well. We study two cases depending
on whether this computation accepts or rejects.

\begin{itemize}
	\item 
First consider the case that $M_j^{u_z}(F_r^{u_z}(0^n))$ definitely rejects. 
As $z$ is even, $0^n\in B_p^{u_z}$ and clearly $0^n\in B_p^w$ for all $w\sqsupseteq u_z$.
This, however, contradicts the assumption that step~$s$ of the construction treating the
task $(i,j,r)$ is not possible.
\item
Next we consider the case that $M_j^{u_z}(F_r^{u_z}(0^n))$ definitely accepts. Then both
$M_i^{u_z}(F_r^{u_z}(0^n))$ and $M_j^{u_z}(F_r^{u_z}(0^n))$ definitely accept. As $u_z$ is $t_{s'-1}$-valid 
by Claim~\ref{claim_90124780576}, we obtain that $u_z$ is even $t$-valid for $t = t_{s'-1}\cup\{(i,j)\mapsto 0\}$.
But then the construction would have chosen $t_{s'} = t$, in contradiction to $t_s(i,j) = -p < 0$.
\end{itemize}
As in both cases we obtain a contradiction, the construction described above is possible.
It remains to show that relative to the final oracle $O$, there exist $\cP^O$-optimal proof systems
for $\oli{K^O}$ and $\cNP^O \cap \ccoNP^O$ does not have $\redm[,O]$-complete problems.

\begin{claim}\label{claim_10470231443}
$\oli{K^O}$ has $\cP^O$-optimal proof systems.
\end{claim}
\begin{proof}
Let $g\in \FP^O$ be an arbitrary proof system for $\oli{K^O}$ and
$a$ be an arbitrary element of $\oli{K^O}$.
Define $f$ to be the following function $\sow\to\sow$:
$$
f(z) = \begin{cases}
g(z')&\text{if $z = 1z'$}\\
y&\text{if $z = 0c(i,x,y)$ for $i\in\N^+$, $x,y\in\N$, and $c(i,x,y)\in O$}\\
a&\text{otherwise}
\end{cases}
$$
By definition, $f\in\FP^O$ and as $g$ is a proof system for $\oli{K^O}$ it holds
$f(\sow)\supseteq \oli{K^O}$. We show $f(\sow)\subseteq \oli{K^O}$. Let $z\in\sow$. Assume $z=0c(i,x,y)$
for $i\in\N^+$, $x,y\in\N$, and $c(i,x,y)\in O$ (otherwise, clearly $f(z)\in \oli{K^O}$).
Let $j>0$
such that $F_j^O$ computes $f$. Let $s$
be large enough such that $w_s$ is defined for $c(i,x,y)$, i.e. $w_s(c(i,x,y)) =1$. 
As $w_s$ is $t_s$-valid, we obtain by V1 that $F_i^{w_s}(x) = y\in \oli{K^{w_s}}$ and by 
Claim~\ref{claim_7378194873} that $F_i^{w_s}(x)$ is defined and $y\in \oli{K^{v}}$ for all $v\sqsupseteq w_s$. 
Then $F_i^O(x)\in \oli{K^O}$. This shows that $f$ is a proof system for $\oli{K^O}$.

It remains to show that each proof system for $\oli{K^O}$ is $\cP^O$-simulated by $f$.
Let $h$ be an arbitrary proof system for $\oli{K^O}$. Then there exists $i>0$ such that
$F_i^O$ computes $h$. By construction,
$t_s(i,i) > 0$, where $s$ is the number of the step that treats the task $i$. 
Consider the following function
$\pi:\sow\to\sow$:
$$
\pi(x) = \begin{cases}
0c(i,x,F_i^O(x))&\text{if $c(i,x,F_i^O(x))\ge t_s(i,i)$}\\
z&\text{if $c(i,x,F_i^O(x)) < t_s(i,i)$ and $z$ is minimal with $f(z) = F_i^O(x)$}
\end{cases}
$$
As $f$ and $F_i^O$ are proof systems for $\oli{K^O}$, for every $x$ there exists $z$ with
$f(z) = F_i^O(x)$. Hence, $\pi$ is total.
Since $t_s(i,i)$ is a constant, $\pi\in \FP\subseteq \FP^O$. It remains to show that
$f(\pi(x)) = F_i^O(x)$ for all $x\in\sow$. If $|x| < t_s(i,i)$, it holds $f(\pi(x)) = F_i^O(x)$. Otherwise,
choose $s'$ large enough such that (i) $t_{s'}(i,i)$ is defined (i.e., $t_{s'}(i,i) = t_s(i,i)$) and 
(ii) $w_{s'}$ is defined for $c(i,x,F_i^{w_{s'}}(x))$. Then, as $w_{s'}$ is $t_{s'}$-valid,
V5 yields that $c(i,x,F_i^{w_{s'}}(x))\in w_{s'}$.
By Claim~\ref{claim_7378194873}, $F_i^{w_{s'}}(x)$ is defined and hence,
$F_i^O(x) = F_i^{w_{s'}}(x)$ as well as $c(i,x,F_i^O(x))\in w_{s'}\subseteq O$.
Hence, $f(\pi(x)) = F_i^O(x)$,
which shows $h=F_i^O\psim[,O] f$.
This completes the proof of Claim~\ref{claim_10470231443}.
\end{proof}

\begin{claim}\label{claim_102470122}
$\cNP^O \cap \ccoNP^O$ does not have $\redm[,O]$-complete problems.
\end{claim}
\begin{proof}
Assume the assertion is wrong, i.e., there exist distinct $i,j\in\N^+$ such
that $L(M_i^O),L(M_j^O) \in \cNP^O$ with $L(M_i^O) = \oli{L(M_j^O)}$ and for every 
$A\in\cNP^O\cap\ccoNP^O$ it holds $A\redm[,O] L(M_i^O)$. 
From $L(M_i^O) = \oli{L(M_j^O)}$ if follows that
for all $s$ there does not exist $z$ such that both
$M_i^{w_s}(z)$ and $M_j^{w_s}(z)$ definitely accept or both
$M_i^{w_s}(z)$ and $M_j^{w_s}(z)$ definitely reject.
Hence, for no $s$ it holds $t_s(i,j) = 0$ and thus, by construction
$t_s(i,j) = -p$ for some $p \in \Podd$ and all sufficiently large $s$.
The latter implies $|O\cap \az^{p^k}| = 1$ for all $k>0$ (cf.\ V3),
which yields $A_p^O = \oli{B_p^O}$, i.e., $A_p^O\in\cNP^O\cap\ccoNP^O$.
Thus, there exists $r$ such that
$A_p^O \redm[,O] L(M_i^O)$ via $F_r^O$.
Let $s$ be the step that treats task $(i,j,r)$.
This step makes sure that there exists $n\in\N^+$ such that
at least one of the following properties holds:
\begin{itemize}
            \item   $0^n\in A_p^{v}$ for all $v\sqsupseteq w_{s}$ 
                    and $M_i^{w_{s}}(F_r^{w_{s}}(0^n))$ definitely rejects.
            \item   $0^n\in B_p^{v}$ for all $v\sqsupseteq w_{s}$ 
                    and $M_j^{w_{s}}(F_r^{w_{s}}(0^n))$ definitely rejects.
\end{itemize}
As $O(q) = w_s(q)$ for all $q$ that $w_s$ is defined for, one of the
following two statements holds.
\begin{itemize}
    \item $0^n \in A_p^O$ and $F_r^O(0^n)\notin L(M_i^O)$.
    \item $0^n \in B_p^O = \oli{A_p^O}$ and $F_r^O(0^n)\notin L(M_j^O) = \oli{L(M_i^O)}$.
\end{itemize}
This is a contradiction to $A_p^O \redm[,O] L(M_i^O)$ via $F_r^O$,
which completes the proof of Claim~\ref{claim_102470122}.
\end{proof}

This finishes the proof of Theorem~\ref{theorem_0917240914}.
\end{proof}

\begin{corollary}
It holds relative to the oracle $O$ of Theorem~\ref{theorem_0917240914}:
\begin{itemize}
	\item $\cNP^O\cap \ccoNP^O$ does not have $\redm[,O]$-complete problems.
    \item Each set complete for $\ccoNP^O$ has $\cP^O$-optimal proof systems.
\end{itemize}
\end{corollary}
\begin{proof}
This follows from Theorem~\ref{theorem_0917240914} and Corollary~\ref{coro_pps_oracle}.
\end{proof}
\bibliographystyle{alpha}

\begin{thebibliography}{{Dos}19b}

\bibitem[CR79]{cr79}
S.~Cook and R.~Reckhow.
\newblock The relative efficiency of propositional proof systems.
\newblock {\em Journal of Symbolic Logic}, 44:36--50, 1979.

\bibitem[DG19]{dg19}
T.~Dose and C.~Gla{\ss}er.
\newblock {NP}-completeness, proof systems, and disjoint {NP}-pairs.
\newblock Technical Report 19-050, Electronic Colloquium on Computational
  Complexity {(ECCC)}, 2019.

\bibitem[Dos19a]{dos19}
T.~Dose.
\newblock P-optimal proof systems for each np-set but no complete disjoint
  np-pairs relative to an oracle.
\newblock In Peter Rossmanith, Pinar Heggernes, and Joost{-}Pieter Katoen,
  editors, {\em 44th International Symposium on Mathematical Foundations of
  Computer Science, {MFCS} 2019, August 26-30, 2019, Aachen, Germany.}, volume
  138 of {\em LIPIcs}, pages 47:1--47:14. Schloss Dagstuhl - Leibniz-Zentrum
  f{\"{u}}r Informatik, 2019.

\bibitem[{Dos}19b]{dos19b}
T.~{Dose}.
\newblock {P-Optimal Proof Systems for Each Set in NP but no Complete Disjoint
  NP-pairs Relative to an Oracle}.
\newblock {\em arXiv e-prints}, pages 1--19, Apr 2019.

\bibitem[GSSZ04]{gssz04}
C.~Gla{\ss}er, A.~L. Selman, S.~Sengupta, and L.~Zhang.
\newblock Disjoint {NP}-pairs.
\newblock {\em SIAM Journal on Computing}, 33(6):1369--1416, 2004.

\bibitem[Kha19]{kha19}
E.~Khaniki.
\newblock {New relations and separations of conjectures about incompleteness in
  the finite domain}.
\newblock {\em arXiv e-prints}, pages 1--25, Apr 2019.

\bibitem[KMT03]{kmt03}
J.~K{\"o}bler, J.~Messner, and J.~Tor\'an.
\newblock Optimal proof systems imply complete sets for promise classes.
\newblock {\em Information and Computation}, 184(1):71--92, 2003.

\bibitem[Pap94]{pap94}
C.~M. Papadimitriou.
\newblock {\em {Computational complexity}}.
\newblock Addison-Wesley, Reading, Massachusetts, 1994.

\bibitem[Pud17]{pud17}
P.~Pudl{\'a}k.
\newblock Incompleteness in the finite domain.
\newblock {\em The Bulletin of Symbolic Logic}, 23(4):405--441, 2017.

\end{thebibliography}



\end{document}